\documentclass{PoS}
\usepackage{psfig,epsfig}
\PoS{PoS(LAT2005)230}

\title{Topological charge renormalization: A test case for 3-loop vacuum
calculations using overlap fermions and Symanzik improved gluons}

\ShortTitle{Topological charge renormalization}

\author{\speaker{Apostolos Skouroupathis}\\
        Univ of Cyprus; Physics Dept\\
        E-mail: \email{php4as01@ucy.ac.cy}}

\author{Haralambos Panagopoulos\\
         Univ of Cyprus; Physics Dept\\
        E-mail: \email{haris@ucy.ac.cy}}

\abstract{We calculate perturbative renormalization properties of the topological 
charge, using the standard lattice discretization given by a product of 
twisted plaquettes. We use the overlap and clover action for fermions, and 
the Symanzik improved gluon action for 4- and 6-link loops.

We compute the multiplicative renormalization of the topological charge density
to one loop; this involves only the gluon part of the action. The power divergent additive renormalization of the topological susceptibility is calculated to 3 loops.}

\FullConference{XXIIIrd International Symposium on Lattice Field
Theory\\
		 25-30 July 2005\\
		 Trinity College, Dublin, Ireland}

\begin{document}

\section{Introduction}
Topological properties of QCD are among those most widely studied on
the lattice. Various methods have been used to this end, involving
renormalization, cooling, fermionic zero modes, geometric definitions,
etc.
In recent years, the advent of fermionic actions, such as the overlap,
which do not violate chirality, has brought a new thrust to the subject.

In this work we compute the renormalization constants which are
necessary in order to extract topological properties, in the ``field
theoretic'' approach, from Monte Carlo simulations using Wilson or
Symanzik improved gluons, and clover or overlap fermions.
We compute the multiplicative renormalization $\mathbf
Z_Q$ of the topological charge density, to 1 loop in perturbation
theory, and the power divergent additive renormalization $\mathbf
M(g^2)$ of the topological susceptibility, to 3 loops.

The main motivations for doing this work are:
a) To enable comparison between different
approaches used in studying topology, so that a coherent picture of
topology in QCD may emerge.
b) To enable studies, in numerical simulations, of
quantities involving the density of topological charge, $q(x)$,
rather than only the integrated charge; this is necessary, e.g., for
studying the spin content of nucleons.
c) As a feasibility study in lattice perturbation
  theory: Indeed, this is the first 3-loop calculation to appear in
  the literature, involving overlap fermions. 

A more extensive write-up of this work can be found in our
Ref.~\cite{SP}.

\section{Computation of $\mathbf Z_Q$}

Our first task is to compute the multiplicative renormalization $Z_Q$
~\cite{CDP}
of the topological charge density $q_{_L}(x)$ to one loop, using the
background field method. We use the standard definition of $q_{_L}$,
given by a product of twisted plaquettes
\begin{equation}
q_{_L}(x)=- {1\over 2^9\pi^2}
\sum^{\pm 4}_{\mu\nu\rho\sigma=\pm 1}
\epsilon_{\mu\nu\rho\sigma} {\rm Tr}
\left[ \Pi_{\mu\nu}(x)\Pi_{\rho\sigma}(x)\right]\;\,
\label{stop}
\end{equation}
($\epsilon_{-\mu,\nu,\rho,\sigma}\equiv
-\epsilon_{\mu,\nu,\rho,\sigma}$ ; in standard notation:
$\Pi_{\mu\nu}(x) = U_\mu(x)\,U_\nu(x+\mu)\,U^\dagger_\mu(x+\nu)\,U^\dagger_\nu(x)$.)

The classical limit of the operator shown in Eq.(\ref{stop}) must be
corrected by including a renormalization function $Z_Q$, which can be
expressed perturbatively as  
\begin{equation}
Z_Q=1+Z_1\cdot{g^2}+\cdots \,\,,\qquad Z_1=Z_{11}\,
\cdotp N_c+Z_{12} / N_c
\label{ZQ}
\end{equation}
We perform a calculation of $Z_1$; this involves only the gluon part of the action.

In the background field method, link variables are decomposed as
\begin{equation}
U_\mu(x)=V_\mu(x)\,U_{c\mu}(x),\qquad V_\mu(x)=e^{igQ_\mu(x)},\qquad
U_{c\mu}(x)=e^{iaB_\mu(x)}
\end{equation}
in terms of links for a quantum field and a classical background
field, respectively.

The diagrams involved in the one-loop calculation of $Z_Q$ are shown
in Figure 1.

The standard Symanzik improved gauge field action, with 4- and 6-link Wilson
loops, is
\begin{eqnarray}
S_G=\frac{2}{g^2} \,\,& \Big[ &c_0 \sum_{\rm plaquette} {\rm Re\, Tr\,}(1-U_{\rm plaquette}) +  c_1 \sum_{\rm rectangle} {\rm Re \, Tr\,}(1- U_{\rm rectangle}) \nonumber \\ 
 & + & c_2 \,\sum_{\rm chair} {\rm Re\, Tr\,}(1- U_{\rm chair})
 +  c_3 \sum_{\rm parallelogram} {\rm Re \,Tr\,}(1- U_{\rm parallelogram})\Big]\,
\label{gluonaction}
\end{eqnarray}
The Symanzik coefficients $c_i$ must satisfy: $c_0 + 8 c_1 + 16 c_2 +
8 c_3 = 1$. 

\begin{figure}
\begin{center}
\epsfig{file=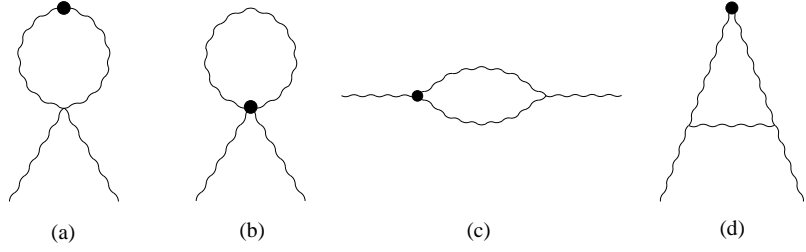,width=.7\textwidth}
\caption{Diagrams contributing to $Z_1$. The
  bullet stands for topological charge vertices.}
\label{fig1}
\end{center}
\end{figure}

Our calculations are performed without assumptions on the values
of the external momenta $p_1,\,p_2$ : This is safest for $q_{_L}$, otherwise one may easily end up with
indeterminate expressions; it entails handling 3-point form
factors~\cite{PanVic} in $D = 4 -2\epsilon$ dimensions, such as: 
\begin{eqnarray}
\bar{C}_{\mu\nu}(a,p_1,p_2)&=&\frac{(k a)^{2
    \epsilon}}{a^0}\int{{\frac{d^Dk}{(2\pi)^D}} {\frac{sin k_{\mu} \,
      sin k_{\nu}}{\hat{k}^2 \,(\widehat{k+a p_1})^2(\widehat{k +a p_1
	+a p_2})^2}}}
\end{eqnarray}

Diagrams (c) and (d) of Figure 1, taken separately, exhibit poles in $\epsilon$
 ( (d) $\propto -1/\epsilon-ln \,\kappa^2
a^2$). These cancel, however, upon summation, as is expected by the
fact that $Q$ does not renormalize in the continuum. 
The calculation of $Z_Q$ is particularly involved in
the present case, involving propagators and vertices from the improved
gluonic action. In particular, the calculation of diagram (d) involves a
summation of $> 1\,000\,000$ different algebraic expressions at
intermediate stages. 

Our results for $Z_Q$ are listed in Table 1.
In all calculations that involve the parameters $c_i$, we choose a
standard set of values, as in Ref.~\cite{HPRSS}. The choice of the sets of parameters correspond to the most popular actions:
The first set corresponds to the plaquette action, the second set
corresponds to the tree-level Symanzik improved action~\cite{Symanzik} 
and the next 6
sets correspond to the tadpole improved L\"uscher-Weisz (TILW)
action~\cite{Alford}
for 6 values of beta:
$\beta =8.60,\,8.45,\,8.30,\,8.20,\,8.10,\,8.00$ .
The last two sets correspond to the Iwasaki~\cite{Iwasaki} and
 DBW2~\cite{Takaishi} actions,
 respectively.

In the case of the plaquette action, our result
agrees with the known result of Ref.~\cite{CDP}.

It is worth noting that the value of $Z_1$ (and of
$e_3$, see below) for the DBW2 action is the smallest one, leading to
a renormalization factor $Z_Q$ closer to 1 (and $M(g^2)$,
Eq.~(\ref{chileq}), closer to 
0). This would single out the DBW2 action as a better candidate for
studies of topology.

\section{Computation of $\mathbf M(g^2)$}

The second task we attend to is the calculation of the additive renormalization of the topological charge susceptibility, which is defined as
\begin{equation}
\chi_{_L}\;=\; \sum_x \langle q_{_L}(x)q_{_L}(0) \rangle 
\label{childef}
\end{equation}
$\chi_{_L}$ develops an unphysical background term which becomes
dominant in the continuum limit
\begin{equation}
\chi_{_L}(g^2)\;=\;a^4 Z_Q(g^2)^2\chi\,+\,M(g^2),\qquad\qquad
M(g^2)=e_3\cdot{g^6}+e_4\cdot{g^8}+\cdots
\label{chileq}
\end{equation}
$M(g^2)$ is the power divergent additive renormalization of
$\chi_{_L}$\,.
We compute the 2-loop coefficient $e_3$\,. This quantity is evaluated
for several
sets of values of the Symanzik improvement coefficients. Figure 2 shows
the diagram contributing to $e_3$\,.
\begin{figure}
\begin{center}
\epsfig{file=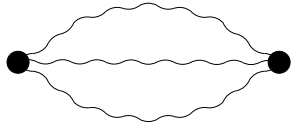,width=.25\textwidth}
\caption{2-loop diagram contributing to
  ${e_3}$. Bullets stand for topological charge vertices.}
\label{fig2}
\end{center}
\end{figure}

The 3-loop term $e_4$ of the expansion of $M(g^2)$ equals
$e_4=e_{4}^g+e_{4}^f$, 
where $e_{4}^f$ stands for the fermionic contribution to $e_4$
($c_{\rm SW}$ is the coefficient in the clover action) 
\begin{equation}
e_{4}^f=N_f(N_{c}^2-1)N_c
\cdot(e_{4,0}+e_{4,1}\,c_{\rm SW}+e_{4,2}\,c_{\rm SW}^2) 
\label{e4f}
\end{equation}
and $e_{4}^g$
is the purely gluonic contribution, expressed as in Ref.~\cite{ACFP}
\begin{equation}
e_{4}^g={1\over{16}}(N_{c}^2-1)(1.735N_{c}^2-10.82+73.83/N_{c}^2) \times{10^{-7}}
\end{equation}
In fact, what we are interested in, is the calculation of the
parameters $e_{4,0}\,,\,e_{4,1}\,,\,e_{4,2}$. This task is performed
using both overlap and clover fermions (clearly, overlap fermions
involve only the parameter $e_{4,0}$). Figure 3 shows the 3-loop
diagrams contributing to the evaluation of $e_4^f$.

\begin{figure}
\begin{center}
\epsfig{file=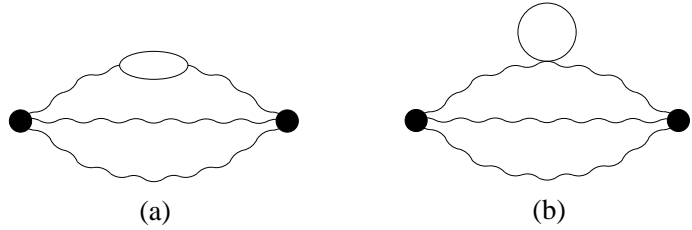,angle=90,width=.6\textwidth}
\caption{Diagrams contributing to $e_4^f$. Wavy (straight) lines
  correspond to gluons (overlap/clover fermions).}
\label{fig3}
\end{center}
\end{figure}

The propagator and vertices of overlap fermions can
be obtained from the following expression for the overlap action,
written in terms of the massless Neuberger-Dirac operator $D_N$ 
~\cite{Neuberger}
\begin{equation}
S_{Overlap}=a^4\sum_{n,\,m}\bar{\Psi}(n)D_N(n,m)\Psi(m),
\qquad\qquad
D_N = \frac{M_0}{a} \left( 1+ \frac{X}{\sqrt{X^\dagger X}}\right)
\label{over}
\end{equation}
$M_0$ is a real parameter corresponding to a negative mass term. $M_0$
must lie in the range $0 < M_0 < 2r$, r being the Wilson perameter (in
our case $r=1$). $X$ is the Wilson-Dirac operator with mass $-M_0$ .

The clover (SW) fermionic action~\cite{SW}, contains an extra term,
parameterized by a coefficient, $c_{\rm SW}$ ; this 
coefficient is treated here as a free parameter.

In performing this calculation, a large effort was
devoted to the creation  of an efficient 3-loop ``integrator'', that
is, a metacode for converting lengthy 3-loop integrands into efficient
code for numerical integration. The output code of the integrator
precalculates a number of time-consuming common
ingredients (Symanzik 
propagator, overlap expressions, etc.), exploits
symmetries of the 
integration region, integrates in parallel over
non-overlapping loops, 
organizes the integrand as an inverse tree for 
optimized evaluation of innermost loops, etc.

Table 1 contains our results for $e_3$
(cf. Eq.(\ref{chileq})) for different gluonic actions. These results,
for the case of the plaquette action, 
agree with older known results (see, e.g., \cite{Christou}).
Figure 4 shows the coefficients $e_{4,0},\ e_{4,1},\
e_{4,2}$ of the clover result for different values of the bare fermion
mass $m$. 
Figure 5 exhibits the dependence of $e_{4}$\,, using
the overlap action, on the parameter $M_0$. 

A complete tabular version of our results on $e_4$\,, for both the
clover and overlap cases, can be found in our longer write-up, Ref.~\cite{SP}.

\begin{table}[ht]
\bigskip
\hspace{-13.5pt}
\begin{tabular}{@{}r@{}lr@{}lr@{}lr@{}lr@{}lr@{}l@{}}
\hline
\hline
\multicolumn{2}{c}{$c_0$}&
\multicolumn{2}{c}{$c_1$}&
\multicolumn{2}{c}{$c_3$}&
\multicolumn{2}{c}{$Z_{11}$}&
\multicolumn{2}{c}{$Z_{12}$}&
\multicolumn{2}{c}{$e_3\times{10^{-7}}$}\\
\hline
\hline
1&.0       & 0&.0      &  0&.0       & -0&.33059398205(2) & 0&.2500000000(1)  & 6&.89791329(1) \\
1&.6666666 &-0&.083333 &  0&.0       & -0&.2512236240(1)  & 0&.183131339233(1)& 3&.1814562840(7) \\
2&.3168064 &-0&.151791 & -0&.0128098 & -0&.20828371039(3) & 0&.147519438874(3)& 1&.8452250005(2) \\
2&.3460240 &-0&.154846 & -0&.0134070 & -0&.20674100461(1) & 0&.146259768983(1)& 1&.8054229585(4) \\
2&.3869776 &-0&.159128 & -0&.0142442 & -0&.20462181183(1) & 0&.144531861677(4)& 1&.7516351593(8) \\
2&.4127840 &-0&.161827 & -0&.0147710 & -0&.20331145580(1) & 0&.143464931830(1)& 1&.7188880608(5) \\
2&.4465400 &-0&.165353 & -0&.0154645 & -0&.20162651307(1) & 0&.142094444611(2)& 1&.6773505020(9) \\
2&.4891712 &-0&.169805 & -0&.0163414 & -0&.19954339172(1) & 0&.140402610424(1)& 1&.626880218(1)  \\
3&.648     &-0&.331    &  0&.0       & -0&.15392854668(1) & 0&.105132852383(2)& 0&.752432061(7)  \\
12&.2688   &-1&.4086   &  0&.0       & -0&.0617777059(4)  & 0&.038277296152(6)& 0&.04881939(4)  \\
\hline
\hline
\end{tabular}
\caption{The values of ${Z_{11}}$ and ${Z_{12}}$ (Eq.(2.2), Figure 1),
  and of ${e_3}$ (Eq.(3.2), Figure 2), with Symanzik improved gluons, 
  for various values of the
  coefficients $c_0,\,c_1,\,c_3$\,.\ \  ($c_2=0$) }
\label{tab1}
\end{table}

\vspace{-2cm}

\begin{center}
\input{./pslatex_files/proc_e4012Vsm.pslatex}

\small{{\bf Figure 4:} Variation of the terms contributing to ${e_{4}^f}$ as a function of $m$}
\end{center}

\vspace{-1cm}

\begin{center}
\input{./pslatex_files/proc_e4VsM0.pslatex}

\small{{\bf Figure 5:} Value of ${e_4}$ as a function $M_0$}
 \end{center}


\vspace{-0.55cm}
\begin{thebibliography}{99}

\bibitem{SP}
A. Skouroupathis and H. Panagopoulos, \emph{Additive and
  multiplicative renormalization of 
      topological charge with improved gluon/fermion actions: A test 
      case for 3-loop vacuum calculations, using overlap or clover fermions},
{\tt hep-lat/0509012}. 

\bibitem{CDP} 
M.~Campostrini, A.~Di Giacomo and  H.~Panagopoulos, \emph{The
  topological susceptibility on the lattice}, \emph{Phys.\ Lett.\ B} {\bf 212} (1988) 206.

\bibitem{PanVic} 
H.~Panagopoulos and E.~Vicari, \emph{The trilinear gluon condensate on
  the lattice}, \emph{Nucl. Phys.} {\bf B332} (1990) 261.

\bibitem{HPRSS}
R.~Horsley et al., \emph{One-loop renormalisation of quark bilinears
  for overlap fermions with improved gauge actions}, \emph{Nucl.Phys.}
{\bf B693} (2004) 3 [Erratum-ibid. {\bf B713} (2005) 601].
 

\bibitem{Symanzik}
K.~Symanzik, \emph{Continuum limit and improved action in lattice
  theories. 1. Principles and $\phi^4$ theory}, \emph{Nucl.\ Phys.} {\bf B226} (1983) 187.

\bibitem{Alford}
M.~G.~Alford, W.~Dimm, G.~P.~Lepage, G.~Hockney and P.~B.~Mackenzie,
\emph{Lattice QCD on small computers}, \emph{Phys.\ Lett. B} {\bf 361} (1995) 87.

\bibitem{Iwasaki}
Y.~Iwasaki, \emph{Renormalization group analysis of lattice theories
  and improved lattice action. 2. Four-dimensional nonabelian SU(N)
  gauge model}, UTHEP-118 (1983).

\bibitem{Takaishi}
T.~Takaishi, \emph{Heavy quark potential and effective actions on
  blocked configurations}, \emph{Phys.\ Rev.} {\bf D54} (1996) 1050.

\bibitem{ACFP} 
B.~All\'es, M.~Campostrini, A.~Feo and H.~Panagopoulos, \emph{Lattice
  perturbation theory by computer algebra: A three loop result for the
  topological susceptibility}, \emph{Nucl. Phys.} {\bf B413} (1994) 553.

\bibitem{Neuberger}
H.~Neuberger, \emph{Exactly massless quarks on the lattice},
\emph{Phys.\ Lett.\ B} {\bf 417} (1998) 141; \emph{More about 
  exactly massless quarks on the lattice}, \emph{Phys.\ Lett.\ B} {\bf 427} (1998) 353.

\bibitem{SW} B.~Sheikholeslami and R.~Wohlert, \emph{Improved
  continuum limit action for QCD with Wilson fermions}, \emph{Nucl. Phys.} {\bf B259}
(1985) 572.

\bibitem{Christou} 
C.~Christou, A.~Di Giacomo, H.~Panagopoulos and E.~Vicari,
\emph{Improved lattice operators: The case of the topological charge density},
\emph{Phys. Rev.} {\bf D53} (1996) 2619.

\end{thebibliography}
\end{document}